\documentstyle[preprint,epsfig,aps]{revtex}
\begin{document}
\newif\iftightenlines\tightenlinesfalse
\tightenlines\tightenlinestrue
\draft
\preprint{\vbox{\hbox{SLAC-PUB-8296}}}

\title{The Paradox of Charmonium Production}

\author{J.\ K.\ Mizukoshi}

\address{\it Stanford Linear Accelerator Center, University of 
Stanford, Stanford, CA 94309, USA}
\date{\today}
\maketitle
\vskip -0.5cm
\begin{abstract} 
The CDF preliminary analysis on polarized charmonium production at moderate
transverse momentum, $p_T \sim 4 - 20$ GeV, severely challenges 
the color octet model (COM), which 
predicts quarkonium to be transversely polarized with increasing 
$p_T$. Based on this data, we analyze the compatibility of  the 
Tevatron and the photoproduction at HERA in the context of the COM. 
Due to the uncertainty on the extraction of non-relativistic QCD (NRQCD)
 matrix elements
and a lack of complete next-to-leading order calculations, one cannot
completely rule out the COM. Nonetheless, both collider experiments
seem to push the input matrix elements to opposite directions, and the 
puzzle of quarkonium polarization remains unsolved.
\end{abstract}

\newpage

\section{Introduction}
The simplest mechanism based on perturbative QCD
to explain quarkonium production, the color singlet model (CSM) 
\cite{csm}, is not able to describe charmonium hadroproduction. 
This model underestimates $J/\psi$ production, both in the 
central \cite{fermi:centralcdf,fermi:centrald0} and forward 
\cite{fermi:forward} rapidity 
regions. The data show that the bound state of heavy-quark pair is 
produced not only in the  color singlet configuration, but there are 
additional states contributing to the final colorless vector meson.
Based on the  non-relativistic QCD model (NRQCD) \cite{nrqcd}, 
quarkonium production  is understood as two-step phenomenon: 
$c \bar{c}$ pair production at perturbative level and the 
subsequent evolution to colorless vector meson through
soft gluon emission at the non-perturbative domain. This argument
 is supported by the fact that the $c \bar{c}$ pair is produced at 
distance $1/m_Q$, $m_Q$ standing for heavy quark mass, which is much smaller 
than $1/\Lambda_{QCD}$, the typical QCD scale for bound-state system. 
According to the color octet model (COM) formulation \cite{com}, a
 generic $S$-wave quarkonium state is  described by the Fock 
state decomposition, schematically given by
\begin{eqnarray}
|\psi_Q\rangle =&& {\cal O}(1)|Q\bar{Q}[^3S_1^{(1)}]\rangle+
{\cal O}(v)|Q\bar{Q}[^3P_J^{(8)}]g\rangle+
{\cal O}(v^2)|Q\bar{Q}[^1S_0^{(8)}]g\rangle+
\nonumber \\
&&{\cal O}(v^2)|Q\bar{Q}[^3S_1^{(8)}]gg\rangle+...
\end{eqnarray}
where $v$ is a typical velocity of the heavy-quark pair. We use 
the usual spectroscopic notation $^{2S+1}L_J$, and the color state is 
indicated by $(1)$ for singlet and $(8)$ for the octet. 

In the first approximation, the  $Q\bar{Q}$ system is produced in a 
color singlet state, which  already has the quantum numbers of 
the physical quarkonium. The octet contributions are suppressed by powers 
of $v$ and $\alpha_s$. The latter is  due to the extra soft 
gluon radiation needed to produce the correct color and/or  
quantum numbers. In principle,
the state $^3P_J^{(8)}$ can produce $\chi_J$ mesons or evolve 
nonperturbatively to a vector meson. 

At the partonic level, the inclusive $\psi$ (generically denoting the 
charmonium $J/\psi$ and $\psi(2S)$) is given by
\begin{equation}
d\hat{\sigma}(a+b \to \psi +X) = \sum_n d\hat{\sigma}_n
(a+b \to c\bar{c}[n] +X) 
\langle {\cal O}_n^{\psi} \rangle,
\label{eq:ab}
\end{equation}
where $c\bar{c}[n]$ stands for the quark-pair in the generic state $n$. 
We denote $\hat{\sigma}_n$ as the cross section for the short 
distance $c$-pair production, which can be calculated perturbatively. 
The matrix elements of the transition 
$c\bar{c}[n] \to \psi$, $\langle {\cal O}_n^{\psi} \rangle$, cannot be 
calculated in the usual perturbation theory. Fortunately, they are assumed 
to be universal, and can be extracted from experiments. 

Of course, in principle one could argue that the dominant long-distance 
matrix element should be $\langle ^3S_1^{(1), \psi} \rangle$,
the $c\bar{c}$ state already with the correct quantum number and 
color of vector meson. However, as already stated, from the Fermilab 
experiments the CSM itself cannot explain the transverse momentum 
$p_T$ of the inclusive reaction $p\bar{p} \to \psi X$. 
The CSM  differential cross section behaves like 
$d\sigma/dp_T \sim 1/p_T^6$, falling much faster than the data. 
The $p_T$ dependence can be fixed combining  the octet states 
$^3S_1^{(8)}$, $^1S_0$, $^3P_J$, which is order of $m_c^3v_c^3$ 
according to NRQCD expansion. Particularly, the $^3S_1^{(8)}$ is fundamental 
for explaining a harder $p_T$ spectrum. If nature favors the vector 
$S$-wave octet state to evolve to a vector meson, there is a strong 
consequence on quarkonium production. Because the $c\bar{c}$ bound-state
is originated from gluon jet, the COM predicts
quarkonium to be transverse polarized on the limit $4m_c^2/p_T^2 \ll 1$ 
\cite{pol:com,pol2:com}. Indeed, it has been shown  that in this limit, the 
charmonium fragmentation function $D_{g \to \psi}$ \cite{pol:frag} could 
be recovered \cite{com}.

The $^3S_1^{(8)}$ plays the major role on the explanation of charmonium 
data at Tevatron, however the same is not true at HERA for $z > 0.2$,
 where $z = p_{\psi} \cdot p_p/p_{\gamma} \cdot p_p$. In this 
kinematic region, the COM predictions for photoproduction 
\cite{com:hera,com2:hera,com:fleming} are dominated by the states 
$^1S_0$ and  $^3P_J$. Moreover, in the Ref.\ \cite{com:fleming}, the authors
show that the polarization of $J/\psi$ produced from
photoproduction depends only on the rate of the matrix elements of 
these states.

Due to the universality, the values of matrix elements extracted 
from CDF data should reproduce HERA data.
However, the COM predicts an excess of events compared with
HERA data \cite{hera:h1,hera:zeus} for $z \to 1$. This discrepancy could be 
explained by the higher-order  QCD effects \cite{hera:high} or 
by the intrinsic 
transverse momentum of the partons \cite{hera:intrin}.
Nonetheless, it seems these tentative solutions cannot reproduce
quite well not only $z$, but other relevant kinematic distributions
\cite{hera:new}.

In the following, we make a quantitative study of the COM in 
the light of the experimental data and we show that is quite difficult to 
accommodate the $J/\psi$ production at Tevatron and  HERA,
simultaneously.  Even with the introduction of higher order QCD 
corrections, the COM will face another challenge: the interpretation 
of charmonium polarization.
The preliminary CDF analysis \cite{polar:preli}
is pointing to unpolarized $\psi$ production, contradicting the COM
predictions.

Our strategy is at follows. We extract independently the 
non-perturbative matrix elements  from both Tevatron and HERA data.  
After determining the solution that could satisfies both data, we show 
that actually it is incompatible to the polarization data.

%
%
\section{hadro and photoproduction in the color octet model}

At the Tevatron, the inclusive $\psi$ production cross section can be 
written as the usual form,
\begin{equation}
d\sigma^{\lambda}_{p\bar{p} \to \psi X}(s) = 
\int f_{a/p}(x_a) f_{b/\bar{p}}
(x_b) \; \hat{\sigma}^{\lambda}_{a b \to \psi X}(\hat{s}),
\label{eq:hadron}
\end{equation}
where the $\hat{\sigma}$ is given by Eq.\ (\ref{eq:ab}) and at 
perturbative level
\begin{equation}
\frac{d\hat{\sigma}^{(\lambda)}_{a b \to c\bar{c} X}[n] }{d\hat{t}}  = 
A_{a b}[n]+
B_{a b}[n][\epsilon(\lambda)\cdot k_a]^2+
C_{a b}[n][\epsilon(\lambda)\cdot k_b]^2+
D_{a b}[n][\epsilon(\lambda)\cdot k_a][\epsilon(\lambda)\cdot k_b],
\label{eq:parton}
\end{equation}
where $k_a$ and $k_b$ are the momenta of the initial partons 
$a$ and $b$ and $\epsilon(\lambda)$ the polarization vector of $\psi$. 
The complete analytic expression can be found in \cite{pol2:com,com2:hera}.
The sum over $\lambda$ yields the unpolarized cross section.
  
Since we want to detect a vector meson, the lowest order at the 
partonic level should be $ 2 \to 2$ process. At the Tevatron, the most 
important contribution comes from the subprocess
\begin{equation}
g \;g \to c\bar{c}[n]\; g,
\end{equation}
although $gq$ and $q\bar{q}$ bring some contribution, especially 
for high $p_T$.

At HERA, there are two types of mechanisms contributing to 
distinct regions of $z$. Direct $\psi$ photoproduction, given by the 
partonic level subprocesses
\begin{eqnarray} 
\gamma^* \;g &\to& c\bar{c}[n]\; g
\\
\gamma^* \;q(\bar{q}) &\to& c\bar{c}[n]\; q(\bar{q}),
\end{eqnarray}
which are important for $z > 0.2$, and by resolved photon mechanism 
through the partonic content of the $\gamma^*$, dominant for small $z$.
All the relevant  analytic expressions are listed in \cite{com2:hera}.

The complete $e p \to \psi X$ can be written as
\begin{equation}
d\sigma_{e p \to e J/\psi X}(s) = \Gamma(Q^2,y) \; 
d {\sigma}_{\gamma^* p \to J/\psi X}(W^2), 
\end{equation}
where
\begin{equation}
\Gamma(Q^2,y) = \frac{\alpha}{2 \pi y Q^2} [1+(1-y^2)^2].
\end{equation}
The $d {\sigma}(W^2)$ can be related to the partonic cross section 
for the resolved photon process according to
\begin{equation}
d {\sigma}_{\gamma^* p \to J/\psi X}(W^2) = \int f_{i/\gamma}(x_i) f_{b/p}
(x_b) \; d\hat{\sigma}_{i b \to \psi X}(\hat{s}).
\end{equation}
For $i=\gamma$, $f_{i/\gamma}(x_i)=\delta(1-x_i)$ reproduces the expression 
for  direct production.

%
\section{Discussions and Conclusions}

We have performed a numerical calculation of the charmonium production 
cross section fixing $m_c = 1.5$ GeV and 
choosing the  renormalization and factorization scale to be 
$\mu = \sqrt{p_{T \psi}^2+m_{\psi}^2}$, where $m_{\psi} = 2 m_c$. 
We make our analysis for three different parton 
distribution functions (PDF's) in the proton; MRS (R2) \cite{mrs}, 
CTEQ 4L \cite{cteq}, and GRV 94 LO \cite{grv}. For the resolved photons, 
we use GRV distribution function \cite{grv:photon}.
As the shape of $c\bar{c}[^1S_0^{(8)}]$ and $c\bar{c}[^3P_J^{(8)}]$ are 
almost identical, we use the usual combination 
$\langle ^1S_0^{(8)} \rangle +\frac{k}{m_c^2} \langle ^3P_J^{(8)} \rangle
\equiv M_k$, fixing $k=3$.

Evidently a careful analysis on $\mu$, as well as $m_c$  dependence 
would bring to better control of the  theoretical uncertainties. 
Overall, even with an uncertainty with  factor two, our 
conclusions still remain valid. Nevertheless, we consider a case with $m_c =
1.3$ GeV for a more complete check.

\begin{figure}
\begin{center}
\mbox{\epsfig{file=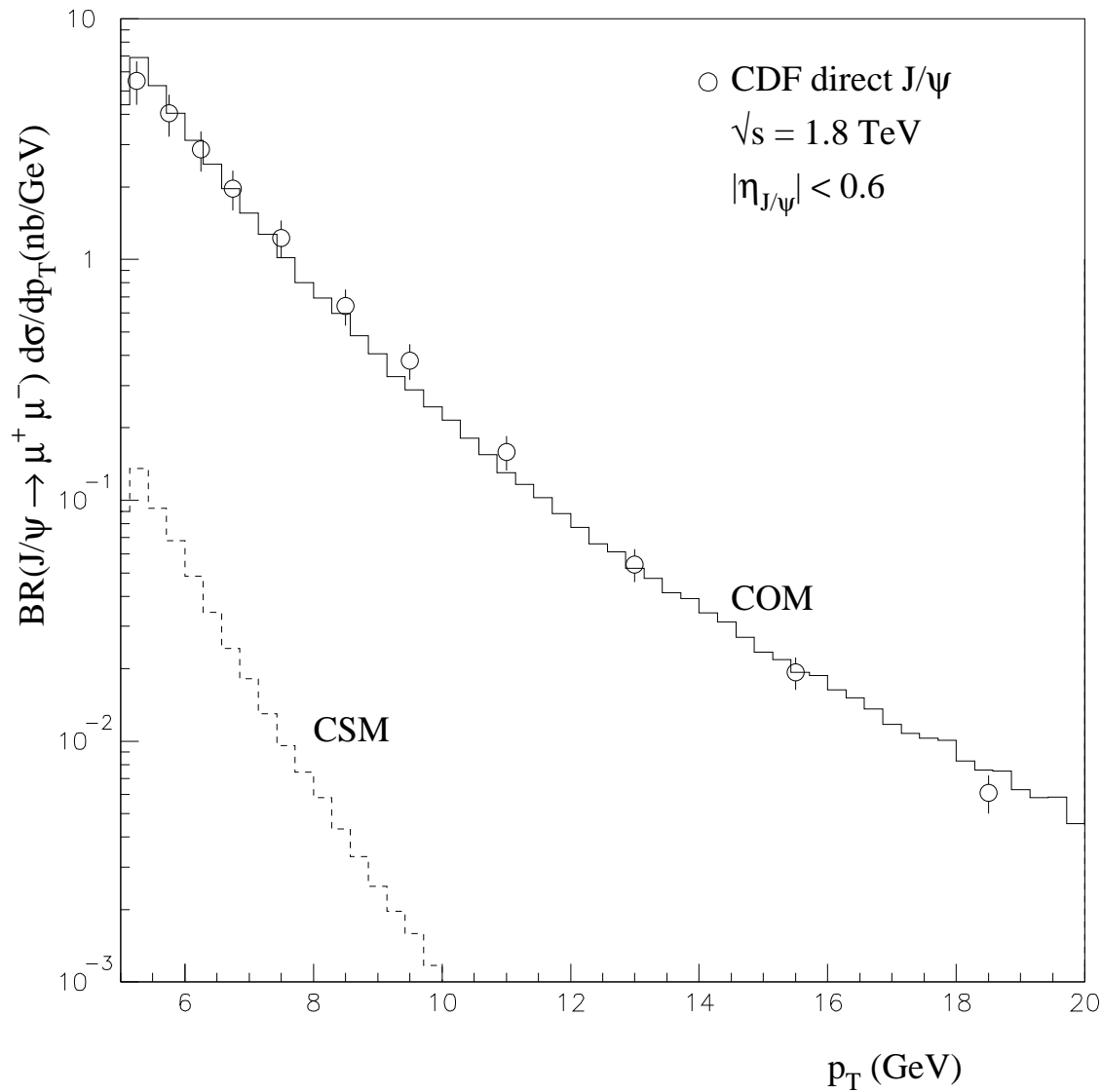,width=1\textwidth}}
\end{center}
\caption{The $p_T$ distribution data (circles) for direct 
forward $J/\psi$ production ($|\eta_{J/\psi}| < 0.6$) from the 
CDF Collaboration at $\sqrt{s} = 1.8$ TeV. The solid curve 
represents the COM prediction after 
choosing the values for the NRQCD matrix elements given in Table 
\ref{tab:matrix} for the CTEQ 4L parton distribution function. 
The dashed  curve shows  the color 
singlet contribution.}
\label{fig:pt_cent}
\end{figure}

As we point out in the introduction, we extract the non-perturbative 
matrix elements independently, for both Tevatron and HERA. 
From the  Fig.\ \ref{fig:pt_cent} we can see that the COM can accommodate 
quite well the CDF central $(|\eta_{\psi}| < 0.6)$  
direct $J/\psi$ production data \cite{fermi:centralcdf}. We should 
emphasize that the same set of matrix elements brings to an extraordinary 
agreement with  the D0 forward $(2.5 < |\eta_{\psi}| < 3.7)$ $J/\psi$ 
production data  \cite{fermi:forward}. However, this is not a surprise,
once we only fit these free universal matrix elements without any 
constraint.

\begin{figure}
\begin{center}
\mbox{\epsfig{file=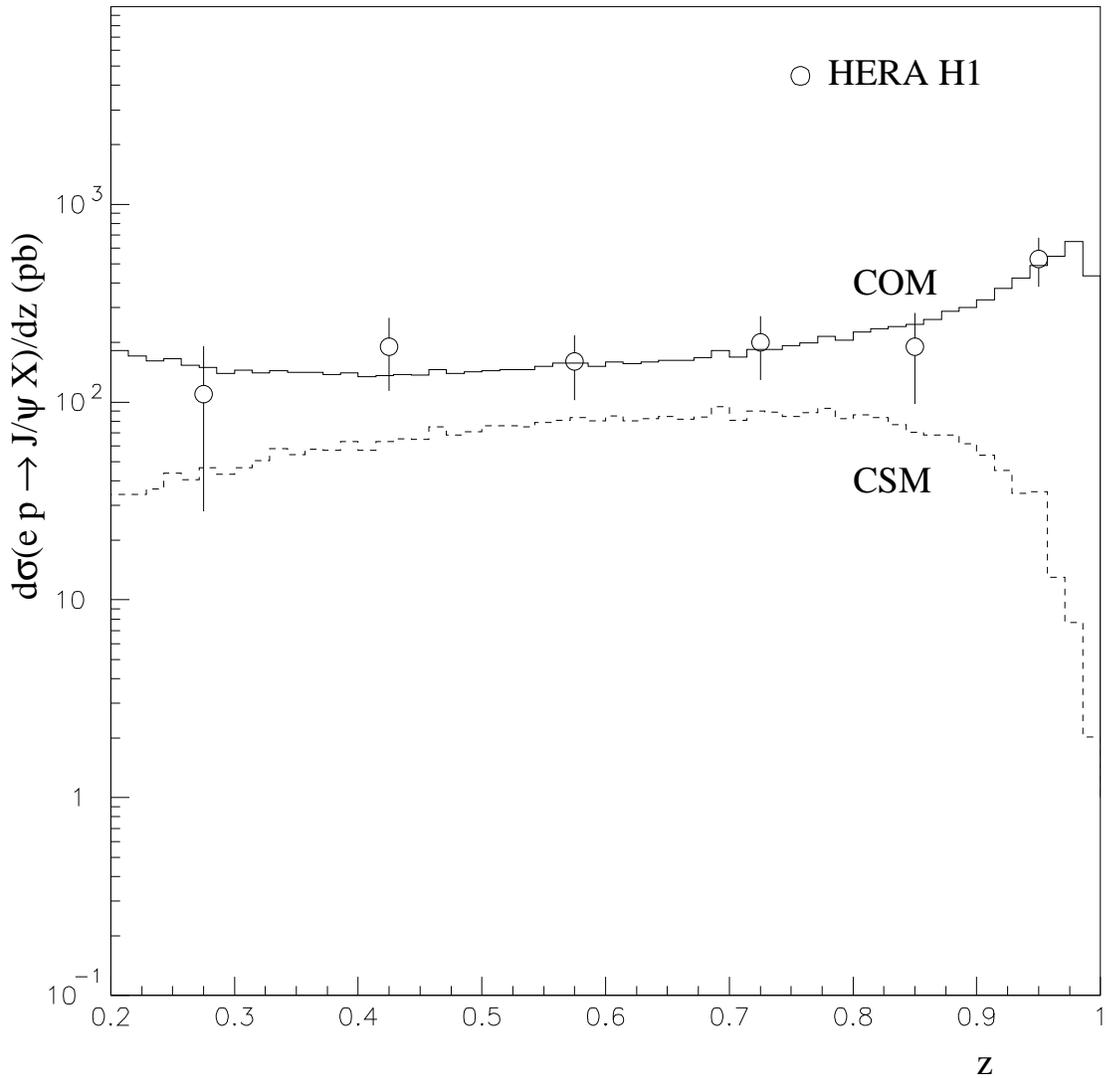,width=1\textwidth}}
\end{center}
\caption{The $z$ distribution  for the inelastic $J/\psi$ production at 
HERA from H1 Collaboration in the kinematic region $ 4 < Q^2 < 80$
GeV$^2$, $p_{T, J/\psi}^2 > 4$ GeV$^2$, $40 < W < 180$ GeV and $z > 0.2$.
The solid curve represents the 
COM prediction after choosing appropriate values for the NRQCD matrix 
elements given in Table \ref{tab:matrix} for CTEQ 4L parton distribution 
function. The dashed curve shows  the color singlet contribution.}
\label{fig:z_h1}
\end{figure}

\begin{figure}
\begin{center}
\mbox{\epsfig{file=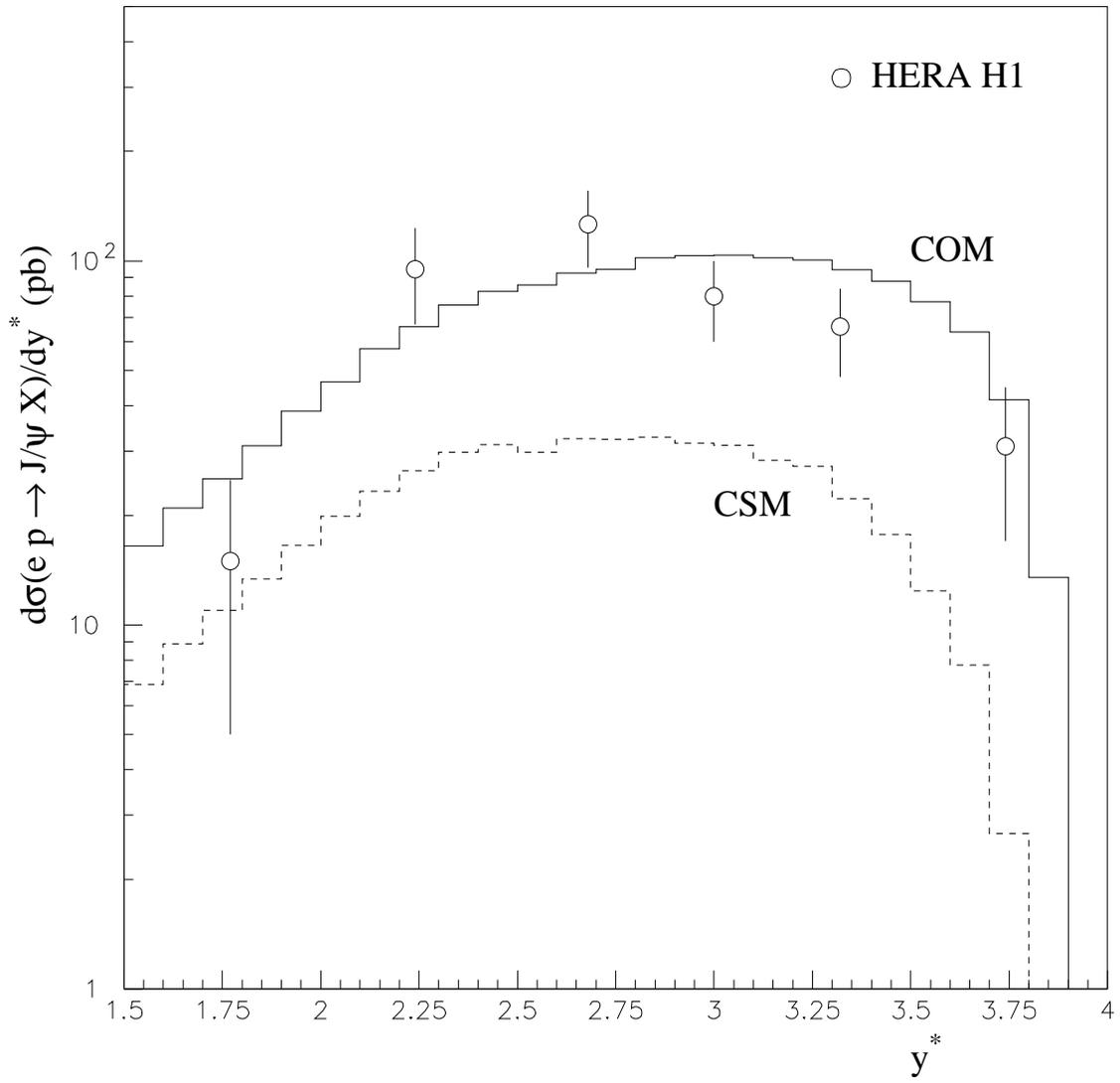,width=1\textwidth}}
\end{center}
\caption{Same as in Fig.\ \ref{fig:z_h1} for the  $y^*$ (the rapidity 
of $J/\psi$ in the $\gamma^*p$ center-of-mass frame) distribution.}
\label{fig:y_h1}
\end{figure}

In the Fig.\ \ref{fig:z_h1} (\ref{fig:y_h1}) we show the $z$ ($y^*$,
the rapidity of $J/\psi$ in the $\gamma^*p$ center-of-mass frame) 
distribution for $J/\psi$ production at 
HERA, and once again, it seems COM can in principle fit well the H1 
data \cite{hera:new}.

In the Table \ref{tab:matrix} we collect the best set of the color octet 
NRQCD matrix elements that fit both data set independently for the three 
PDF's we are considering. For the color singlet contribution, we have used 
$\langle ^3S_1^{(1)} \rangle $ = 1.2 GeV$^3$, following \cite{com}. 
Since  MRS (R2) is calculated at next to leading order, 
it is not surprising that  it gives a different result compared to the 
leading order (LO)  ones. Our numbers confirm the early results 
pointing out that at LO the COM has trouble explaining both  data 
simultaneously. 

As we mention in the Introduction, this anomaly may be cured by adding 
corrections due to the  intrinsic transverse momentum of the partons 
\cite{hera:intrin} or higher-order (HO)  QCD effects \cite{hera:high}. For 
small values of $p_T$, the multiple-gluon radiations 
from the initial and the final state at the Tevatron  become sizeable. 
In the \cite{tevatron:correc} these
corrections were estimated by Monte Carlo simulation using PYTHIA \cite{pythia}
and the HO QCD could be parameterized  as a K factor dependent on $p_T$ 
\cite{hera:high}. In fact, such corrections produce
\begin{eqnarray*}
\langle ^3S_1^{(8)} \rangle &=& (0.47 \pm 0.09) \;\times 10^{-2} \mbox{GeV}^3
\\
M_3 &=& (0.63 \pm 0.34) \;\times 10^{-2} \mbox{GeV}^3
\end{eqnarray*}
for MRS (R2) PDF. This lower value of $M_3$ brings to a better 
agreement with the one extracted from HERA experiments.

\begin{figure}
\begin{center}
\mbox{\epsfig{file=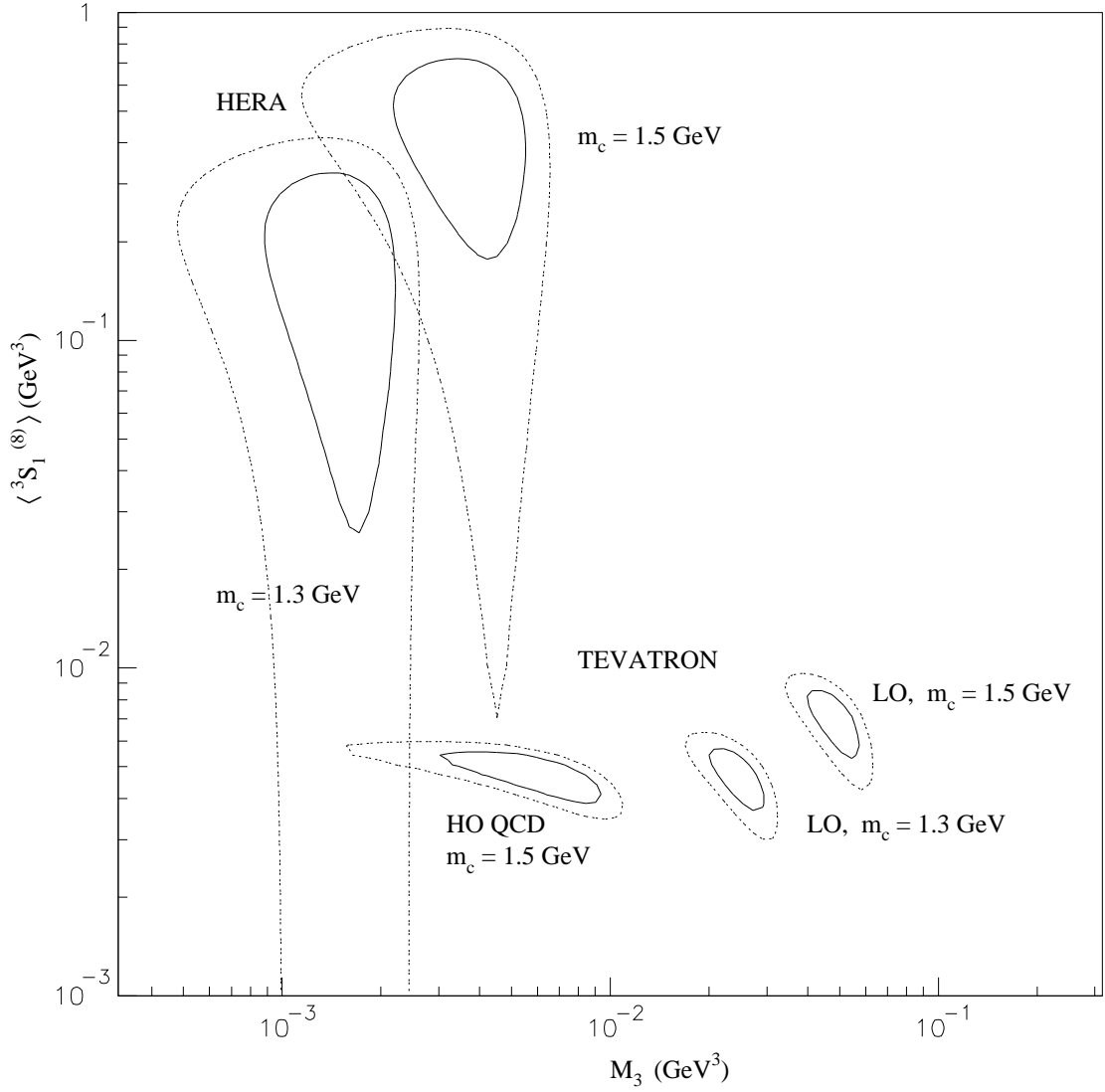,width=1\textwidth}}
\end{center}
\caption{Parameter space for the Color Octet NRQCD matrix elements. 
The bounds on $\langle ^3S_1^{(8)} \rangle $  and $M_3$ for Tevatron and HERA
are displaced at 68\% C.L. (solid lines) and 95\% C.L. (dashed lines). The 
results are for MRS (R2) parton distribution function.}
\label{fig:resumo}
\end{figure}

In the Fig.\ \ref{fig:resumo} we display a parameter space for the 
color-octet  NRQCD matrix elements. Although at 68\% C.L. we still observe 
discrepancy between the bound for Tevatron HO QCD corrections and HERA, 
this picture changes dramatically at 95\% C.L. We see that HERA favors 
much higher values for $\langle ^3S_1^{(8)} \rangle$ than Tevatron, 
however they are not severely constrained.  The reason is that the  state
$^3S_1^{(8)}$ contributes only through resolved photon processes 
in a region where $z < 0.4$ , roughly speaking. In this intermediate 
region, the color-singlet contribution has a major role. Actually, this 
is clear if we consider a different $c$-quark mass value. For $m_c = 1.3$ GeV,
the state $^3S_1^{(1)}$ has a bigger contribution, much closer
to the experimental data. This means the CSM itself could describe the HERA 
data reasonable well, except in the region $z \to 1$.

The main conclusion from the Fig.\ \ref{fig:resumo} is that the
introduction of HO QCD corrections brings a match between
COM predictions at HERA and Tevatron, as already pointed 
in \cite{hera:high}.

With the extraction of the NRQCD matrix elements,
we are now in the position to discuss the implication  of these results 
on the charmonium polarization predicted by the COM. 

The quarkonium polarization can be measured from the angular dependence 
on $\psi \to \mu^+ \mu^-$,
\begin{equation}
\frac{d \Gamma}{d \cos \theta} \propto 1+ \alpha \cos^2 \theta,
\end{equation}
with $\alpha = (1-3\xi)/(1+\xi), \; \xi \equiv
d\sigma^{\lambda=0}/ \sum_{\lambda} d\sigma^{\lambda}$.

\begin{figure}
\begin{center}
\mbox{\epsfig{file=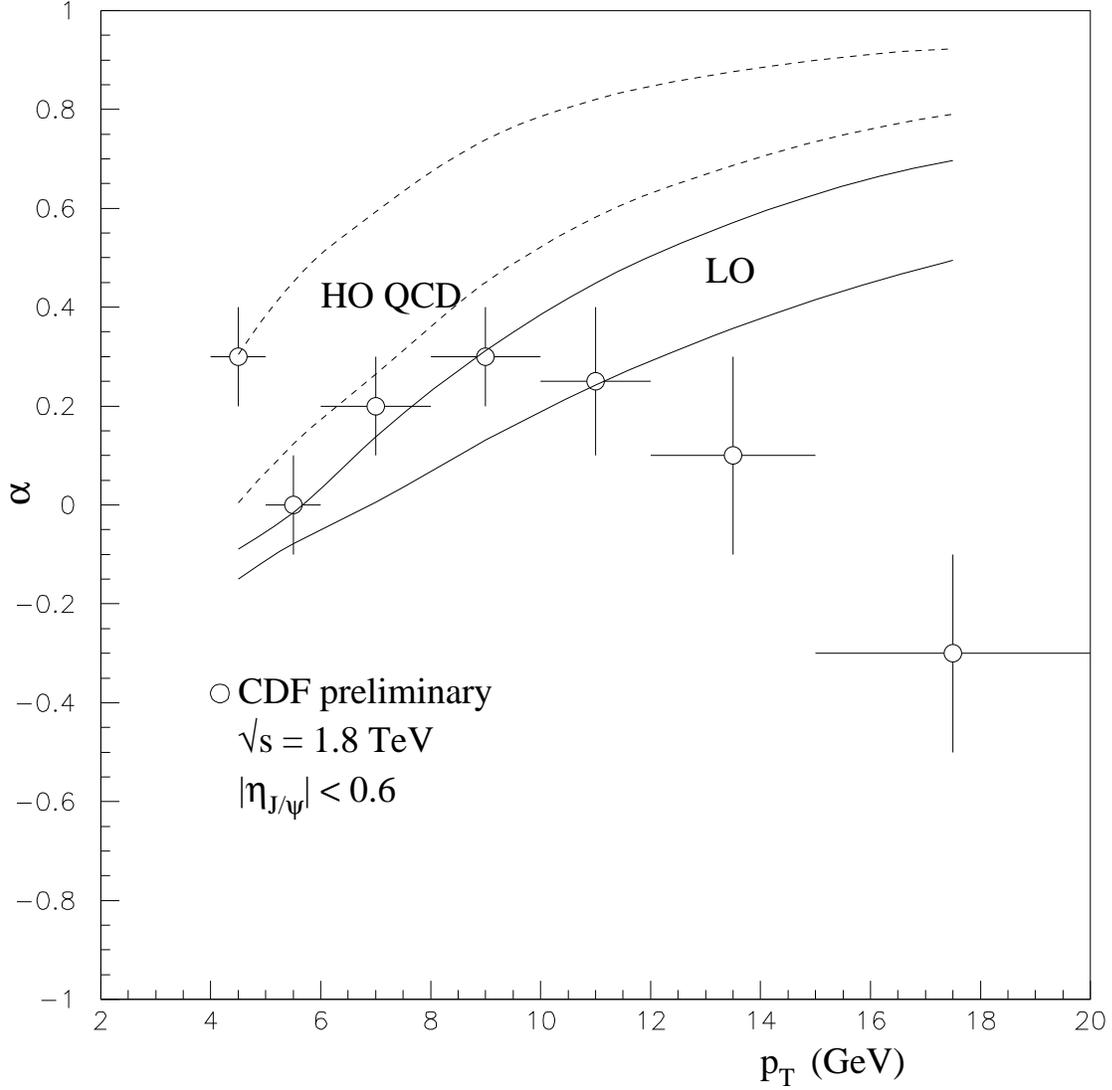,width=1\textwidth}}
\end{center}
\caption{The polarization parameter $\alpha$ as a function of  $p_T$ for
the inclusive prompt $J/\psi$ production at the Tevatron. The bounds
at LO (solid lines) and HO QCD (dashed lines) are based on 68\% C.L. 
including only errors due to the experimental data from CDF preliminary 
analysis  for $|y_{J/\psi}| < 0.6$. The results are for MRS (R2) 
parton distribution function.}
\label{fig:alpha}
\end{figure}

From the expressions in \cite{com2:hera}, it is possible to calculate
the quarkonium cross section for each polarization through the Eqs.\ 
(\ref{eq:hadron}) and (\ref{eq:parton}). Writing the polarization vector 
of quarkonium in the recoil ($s$-channel helicity) frame 
\cite{polar:frame}, we found the $p_T$ dependence on $\alpha$, as displaced
in the Fig.\ \ref{fig:alpha}.  The polarizations were calculated for
seven $p_T$ bins, specified in \cite{polar:preli}. 

At this point, we should be careful in comparing our results with the 
CDF preliminary analysis, once the data contain feed-down from $\chi_c$, 
which contributes to $\sim$ 35\%, and $\psi(2S)$ decay to $J/\psi$.
 
In order to avoid these extra contributions to the prompt charmonium 
production, we also performed analysis on $\psi(2S)$ productions, which do 
not receive feed-down contributions, neither decay of the higher states.
Although from the theoretical point of view the $\psi(2S)$ state is simpler
to analyze, the available data \cite{fermi:centralcdf} are more 
limited statistically. 

\begin{figure}
\begin{center}
\mbox{\epsfig{file=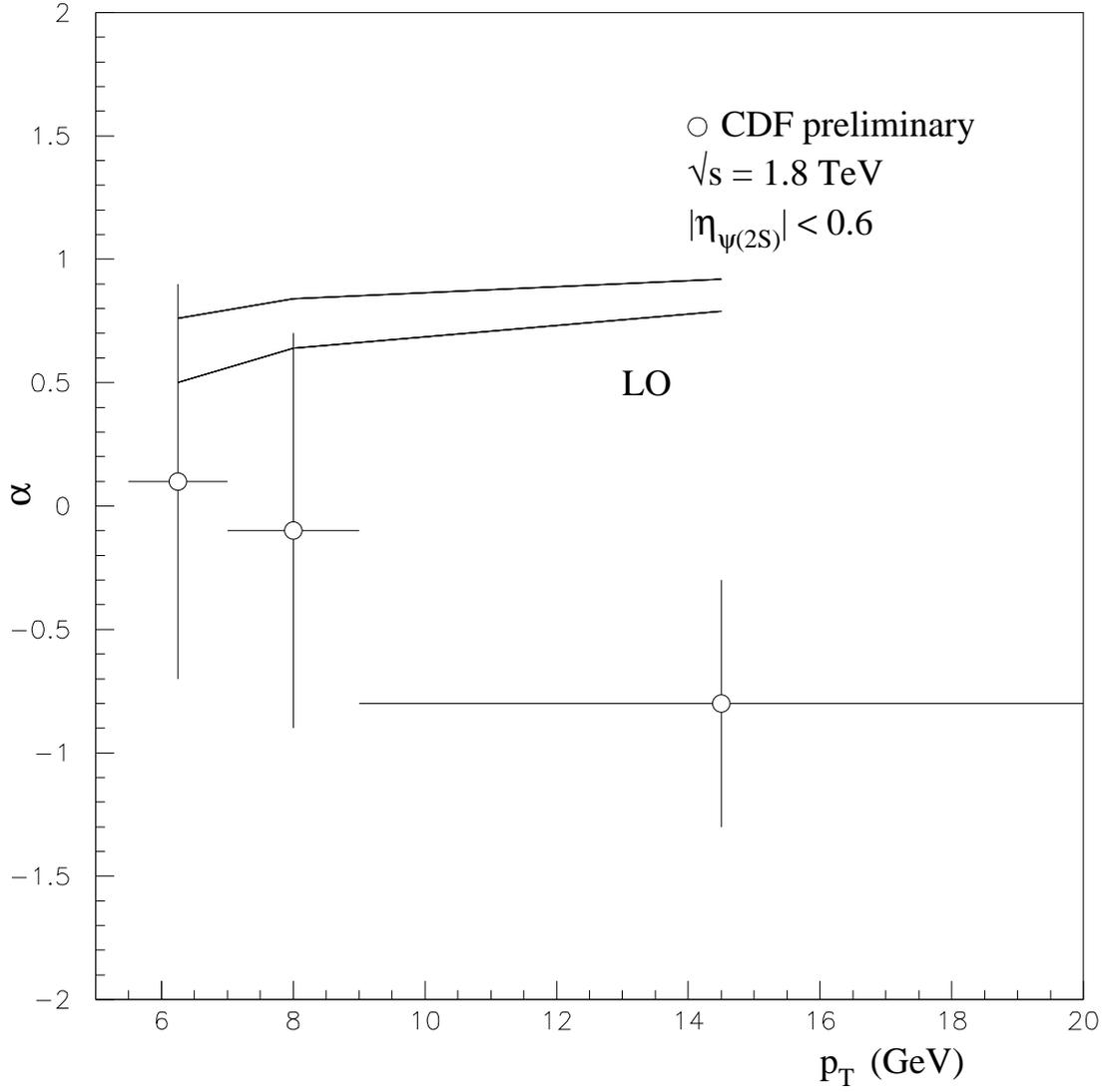,width=1\textwidth}}
\end{center}
\caption{Same as Fig.\ \ref{fig:alpha} for the $\psi(2S)$ 
production at the Tevatron.}
\label{fig:alpha_psip}
\end{figure}

We extracted  the NRQCD matrix elements in a similar way we have done for 
$J/\psi$. For the MRS (R2) PDF, we found $\langle ^3S_1^{(8)} \rangle =
(0.14 \pm 0.03) \times 10^{-2} \;\rm{GeV}^3$ and very small value for
$M_3$, compatible with zero. Following \cite{polar:preli}, we 
calculated the $\psi(2S)$ polarization
for three $p_T$ bins, and once again the charmonium was found to be
transverse polarized, as we can see from the Fig.\ \ref{fig:alpha_psip}. 

The HO QCD corrections, that worked well to solve the Tevatron/HERA 
discrepancy, actually worsen the already poor LO predictions, as the Fig.\ 
\ref{fig:alpha} indicates.

In order to satisfy polarization data, the $\langle ^3S_1^{(8)} \rangle $
contribution must vanish, since the $M_3$ brings to a almost unpolarized 
$\psi$ production. Of course, there is a penalty doing just adjustment
by hand. The nice fit, {\it e.g.},  Fig.\ \ref{fig:pt_cent} is no 
longer held. Besides, the $M_3$ value should be increased in order to have a 
better fit. From the Fig.\ \ref{fig:resumo}, we see that this scenario 
will be disastrous if we compare with HERA bounds.

Although there are strong evidences that COM is not working well, before 
ruling out this model, we should investigate the possible solutions
to solve the paradox of charmonium production:
\begin{itemize}

\item The complete QCD higher order corrections, which is not available so 
far, and contributions from higher $c\bar{c}$ states 
could in principle bring to a drastic change in the scenario.

\item The emitted gluons, in order to  produce the physical quarkonium, 
are not so soft. Therefore, the polarization of the $c\bar{c}$ system 
is not conserved during the evolution to non-perturbative regime.

\end{itemize}

On the other hand, although it is a strong statement, we could argue that
the evolution of $Q\bar{Q}$ system to a physical vector boson is not 
well understood; the splitting  between perturbative and 
non-perturbative regimes cannot be done trivially. This means that 
NRQCD is not appropriate to describe quarkonium production.  Actually, 
if we remember that for the charmonium the perturbative 
expansion is based on ${\cal O} (m_c v_c)$, maybe $m_c$ is not 
sufficiently small to allow such expansion. A closer examination on
bottomonium states will be crucial to check if this state is held or not.

Although is still early to make any strong conclusions, it seems that
the COM is once again in trouble. At least in leading order cannot 
explain simultaneously the Tevatron and HERA data. The existing
solution,  the implementation of HO QCD corrections, worsen the strong
prediction of this model, the production of transverse polarized quarkonium.


\acknowledgments

We thank S.\ J.\ Brodsky, J.\ C.\  Rathsman, and  A.\ Petrov for 
useful discussions. This work was supported by Funda\c{c}\~ ao de Amparo 
\`a Pesquisa do  Estado de S\~ ao Paulo (FAPESP), 
Brazil, and by U.\ S.\ Department  of Energy under contract DE-AC03-76SF00515.

\medskip



\begin{table}
\begin{tabular}{||c|c|c|c||}
            & MRS (R2) & CTEQ 4L  &  GRV 94 LO  \\
\hline
$\langle ^3S_1^{(8)} \rangle $   &  0.70 $\pm \;0.17$ (45 $\pm \;29$)  
& 0.54 $\pm \;0.12 $ (25 $\pm \;22$) & 0.57 $\pm \; 0.12$  (24 $\pm \;22$) \\
\hline
$\langle ^1S_0^{(8)} \rangle +\frac{3}{m_c^2} \langle ^3P_J^{(8)} \rangle$ 
& 4.85 $\pm \;0.95$ (0.39 $\pm \;0.18 $)  & 2.28 $\pm \;0.55$  
(0.29 $\pm \;0.14$ ) & 2.07 $\pm \; 0.53$ (0.30 $\pm \; 0.14$) \\
\end{tabular}
\medskip\medskip
\caption{Leading-order Color-octet NRQCD matrix elements in 
units of $10^{-2}\; \mbox{GeV}^3$ for
direct $J/\psi$ production at Tevatron (HERA).}
\label{tab:matrix}
\end{table}


%
%
%
%
%
%
%
%
%

\end{document}